\documentclass[%
 reprint,
superscriptaddress,
nobibnotes,
 amsmath,amssymb,
 aps,prd,
onecolumn
]{revtex4-2}

\usepackage{graphicx}
\usepackage{dcolumn}
\usepackage{bm}
\usepackage[mathlines]{lineno}

\usepackage{comment}
\usepackage{hyperref}
\usepackage{cleveref}
\usepackage[usenames]{color}

\newcommand{\jcap}{J. Cosmol. Astropart. Phys.}
\newcommand{\mnras}{Monthly Notices of the Royal Astronomical Society}

\newcommand{\araa}{Annual Review of Astronomy and Astrophysics}

\newcommand{\aap}{Astronomy and Astrophysics}

\newcommand{\physrep}{Physics Reports}
\newcommand{\aj}{the Astronomical Journal}
\newcommand{\apss}{Astrophysics and Space Science}

\usepackage[dvipsnames]{xcolor}  

\usepackage[normalem]{ulem} 

\begin{document}

\preprint{APS/123-QED}

\preprint{APS/123-QED}
\title{The role of LRG1 and LRG2's monopole in inferring the DESI 2024 BAO cosmology}

\author{Zhengyi Wang}
\affiliation{Institute for Frontier in Astronomy and Astrophysics, Beijing Normal University, Beijing 102206, China}
\affiliation{
Department of Astronomy, Beijing Normal University, Beijing 100875, China}

\author{Shijie Lin}%
\affiliation{Institute for Frontier in Astronomy and Astrophysics, Beijing Normal University, Beijing 102206, China}
\affiliation{
Department of Astronomy, Beijing Normal University, Beijing 100875, China}%

\author{Zhejie Ding}
\affiliation{
Department of Astronomy, School of Physics and Astronomy, Shanghai Jiao Tong University, Shanghai 200240, China}
\affiliation{Shanghai Key Laboratory for Particle Physics and Cosmology, Shanghai 200240, China
}%

\author{Bin Hu}
\email{bhu@bnu.edu.cn}
\affiliation{Institute for Frontier in Astronomy and Astrophysics, Beijing Normal University, Beijing 102206, China}
\affiliation{
Department of Astronomy, Beijing Normal University, Beijing 100875, China}


\begin{abstract}
The Dark Energy Spectroscopic Instrument (DESI) collaboration recently released its first year of data (DR1) on baryon acoustic oscillations (BAO) in galaxy, quasar, and Lyman-$\alpha$ forest tracers. When combined with CMB and SNIa data, DESI BAO results suggest potential thawing behavior in dark energy. Cosmological analyses utilize comoving distances along ($D_H$) and perpendicular to ($D_M$) the line of sight. Notably, there are $1\sim2\sigma$ deviations in $D_M$ and $D_H$ from Planck cosmology values in the luminous red galaxies (LRG) bins LRG1 and LRG2.This study examines the role of LRG1 and LRG2 in diverging DESI 2024 BAO cosmology from Planck cosmology. We use angle-averaged distance $D_V$ and the ratio $F_{\rm AP}=D_M/D_H$, which are more directly related to the measured monopole and quadrupole components of the galaxy power spectrum or correlation function, instead of the officially adopted $D_M$ and $D_H$. This transformation aims to isolate the influence of monopoles in LRG1 and LRG2 on deviations from $w=-1$.
Our findings indicate that removing the $D_V$ data point in LRG2 aligns DESI + CMB + SNIa data compilation with $w=-1$ within a $2\sigma$ contour and reduces the $H_0$ discrepancy from the Planck 2018 results from $0.63\sigma$ to $0.31\sigma$. Similarly, excluding the $D_V$ data point from LRG1 shifts the $w_0/w_a$ contour toward $w=-1$, although no intersection occurs. This highlights the preference of both LRG1 and LRG2 BAO monopole components for the thawing dark energy model, with LRG2 showing a stronger preference. We provide the $D_V$ and $F_{\rm AP}$ data and their covariance alongside this paper.
\end{abstract}


\maketitle

\section{\label{sec:intro}Introduction}

Dark energy was introduced to explain the discovery of the universe's accelerated expansion \citep{2013PhR...530...87W,2008ARA&A..46..385F}. Using measurements from standard rulers such as baryon acoustic oscillations (BAO), researchers can constrain whether dark energy behaves like a cosmological constant or exhibits physical dynamics. The equation of state of dark energy can be phenomenologically parameterized into the form of $w(z) = w_{0} + w_{a}(1-a)$ \citep{Chevallier:2000qy,Linder:2002et}, where $a$ is the scale factor. 
The BAO signal was first detected by surveys such as the Sloan Digital Sky Survey (SDSS) \citep{2005ApJ...633..560E} and the 2dF Galaxy Redshift Survey (2dFGRS) \citep{2005MNRAS.362..505C}. To achieve more precise distance determinations at the percent level, a new generation of galaxy clustering surveys, including the Six-degree Field Galaxy Survey (6dFGS) \citep{2011MNRAS.416.3017B}, the Baryon Oscillation Spectroscopic Survey (BOSS) \citep{2017MNRAS.470.2617A}, the extended Baryon Oscillation Spectroscopic Survey (eBOSS) \citep{2021PhRvD.103h3533A}, and the WiggleZ Survey \citep{2012MNRAS.425..405B}, were conducted. These surveys have improved BAO distance determinations to an accuracy of $1-2\%$ at $z<1$ \citep{2014MNRAS.441...24A,2017MNRAS.464.3409B,2017MNRAS.469.3762W,2021MNRAS.500..736B}, extended the BAO measurement at $z>1$ \citep{Bautista2017,Ata2018,Hou2021,2020ApJ...901..153D}, and found that dark energy was consistent with the cosmological constant \citep[e.g.][]{eBOSS:2020yzd}.

Recently, the Dark Energy Spectroscopic Instrument (DESI) survey \citep{DESI2016a.Science,DESI2022.KP1.Instr,DESI2023a.KP1.SV,DESI2023b.KP1.EDR}, one of the Stage IV surveys \citep{Euclid2011,WFIST2015,LSST2009}, presented Data Release 1 (DR1) of BAO measurements from galaxy, quasar and Lyman-$\alpha$ forest tracers~\citep{DESI2024.III.KP4,DESI2024.IV.KP6} and the cosmological results derived from these measurements \citep{DESI2024.VI.KP7A}. For the cosmological constraint, the DESI collaboration using the distance measurements (the comoving angular diameter distance, $D_{M}$ and the comoving Hubble distance, $D_{H}$) alone and in combination with other cosmological probes such as the Big Bang Nucleosynthesis (BBN), the Cosmic Microwave Background (CMB), and Type Ia supernovae (SNIa), DESI provided cosmological inference results for various cosmological models, including the standard $\Lambda$CDM model, the $w$CDM model, and the $w_0w_a$CDM model. 
In particular, by fitting the $w_0w_a$CDM model to the DESI $+$ CMB combination, the collaboration reported a significance of 2.6$\sigma$ hint for the thawing dark energy model \citep{Caldwell:2005tm}. 
This significance persisted or increased when Pantheon+ \citep{Pantheon+2022}, Union3 \citep{Union3_2023}, or DES-SN5YR SNIa \citep{DES-SN5YR_2024} were included, leading to results that were discrepant with the $\Lambda$CDM model at the levels of 2.5$\sigma$, 3.5$\sigma$, or 3.9$\sigma$, respectively.

The discrepancies with the $\Lambda$CDM model observed in DESI BAO data have prompted extensive discussions on dark energy. Several studies have sought to constrain various cosmological physics using the new DESI BAO data \citep{2024arXiv240406796W,2024arXiv240408056C,tada2024quintessential,2024arXiv240407070L,2024arXiv240414341B,2024arXiv240415232G,2024arXiv240415220A,Wang:2024dka,qu2024accelerated,Yang:2024kdo,escamillarivera2024ft}, while others have delved into the reasons behind the observed discrepancies \citep{yin2024cosmic,2024arXiv240408633C,2024arXiv240412068C,2024arXiv240413833W,Park:2024jns}. We notice that the DESI distance $D_{H}$ measurement in LRG1 ($z_{\rm eff}=0.510$) and $D_M$ measurement in LRG2 ($z_{\rm eff}=0.706$) exhibit significant deviations from the fiducial Planck cosmology. Some studies have opted to reanalyze cosmological constraints without considering LRG1 \citep{2024arXiv240408633C,2024arXiv240412068C,2024arXiv240413833W}. For instance, Wang \citep{2024arXiv240413833W} shows that by 
combining DESI, Planck and Pantheon+ data, the LRG1 has little impact on the constraint on $w_0, w_a$ parameter estimation. 
Similar conclusion was made by the DESI collaboration \citep{DESI2024.VI.KP7A}. Comparing the Figure 12. and Figure 6. of Ref. \citep{DESI2024.VI.KP7A}, we can see that with or without DESI LRG1 does not change the conclusion. 
However, we should notice that the measurements of $D_{M}$ and $D_{H}$ in LRG1 and LRG2 exhibited a high level of correlation. The corresponding correlation coefficients of $D_{M}$ and $D_{H}$ are $\sim -0.4$, as shown in Table 18 of Ref. \citep{DESI2024.III.KP4}. 

The DESI BAO distance measurements are obtained by fitting ten-ish parameters to the galaxy clustering correlation function or power spectrum. The two most important fitting parameters are the isotropic BAO dilation $\alpha_{\rm iso}$ and the anisotropic BAO dilation $\alpha_{\rm AP}$. 
We reference Table 5 in Ref. \citep{DESI2024.III.KP4} for the detailed list of these fitting parameters.
In the Fourier-space fitting framework, the galaxy power spectrum, $P(k,\mu)$, is written as a function of the wave number $k$ and (cosine) angle $\mu$. The coordinates $(k,\;\mu)$ in the true cosmology are related with those in the fiducial cosmology by 
\begin{eqnarray}
\label{eq:alphaiso}
    k_{\rm true}&=&\frac{\alpha_{\rm AP}^{1/3}}{\alpha_{\rm iso}}\left[1+\mu^2_{\rm fid}\left(\frac{1}{\alpha^2_{\rm AP}}-1\right)\right]^{1/2}k_{\rm fid}\;,\nonumber\\
    \mu_{\rm true}&=&\frac{\mu_{\rm fid}}{\alpha_{\rm AP}}\left[1+\mu^2_{\rm fid}\left(\frac{1}{\alpha^2_{\rm AP}}-1\right)\right]^{-1/2}\;,
\end{eqnarray}
where the subscript ``true'' and ``fid'' denote the quantities in the true and fiducial cosmologies, respectively. 
From Eq. (\ref{eq:alphaiso}), we can see that $\alpha_{\rm iso}$ merely rescale the wave number but not the separation angle. It indicates that the constraining power on $\alpha_{\rm iso}$ only comes from the monopole component of the galaxy power spectrum, namely the isotropic component. 
Meanwhile, $\alpha_{\rm AP}$ enters simultaneously into the rescaling of the wave number and separation angle. Hence, it is involved both in the monopole and quadrupole components.  
Via a global fitting to the galaxy power spectrum or correlation function, one can get the posterior distribution of $\alpha_{\rm iso}$ and $\alpha_{\rm AP}$. 
The final angle-averaged distance, $D_V(z)$, and Alcock-Paczynski (AP) factor, $F_{\rm AP}(z)$, are given by
\begin{eqnarray}
    \frac{D_V(z)}{r_s}&=&\alpha_{\rm iso}\frac{D^{\rm fid}_V(z)}{r_s^{\rm fid}}\;,\\
    F_{\rm AP}(z)&=&\frac{F_{\rm AP}^{\rm fid}(z)}{\alpha_{\rm AP}}\;. 
\end{eqnarray}
From these operations, one can see that the angle-averaged distance, $D_V(z)$, is constrained only by the monopole component of the galaxy power spectrum, but the AP factor, $F_{\rm AP}(z)$, enters into both the monopole and quadrupole components. 
From Figure 7 of Ref. \citep{DESI2024.III.KP4}, one can see that the data quality of the monopole is much better than the quadrupole. Hence, instead of using the officially adopted $D_M$ and $D_H$ observables, we switch to $D_V$ and $F_{\rm AP}$ variables for studying the influences of the monopole and quadrupole to the cosmological implications. 
Notably, the $D_V$ data points in LRG1 and LRG2 deviate from the fiducial Planck cosmology at the level of $1\sim 2\sigma$, as shown in Figure 15 of Ref. \citep{DESI2024.III.KP4}. It motivates us to re-analyze the LRG1 and LRG2 by using $D_{V}$ and $F_{\rm AP}$ instead. 
The rest of this paper are organised in the following ways. In Section \ref{Data}, we present the data and methodology. In Section \ref{results}, we show our validation and the cosmological analysis without $D_{V}$ or $F_{\rm AP}$ or both in LRG1 and LRG2, respectively. Finally, we briefly discuss our findings. 

\begin{figure}[h]
\includegraphics[height=0.45\textwidth]{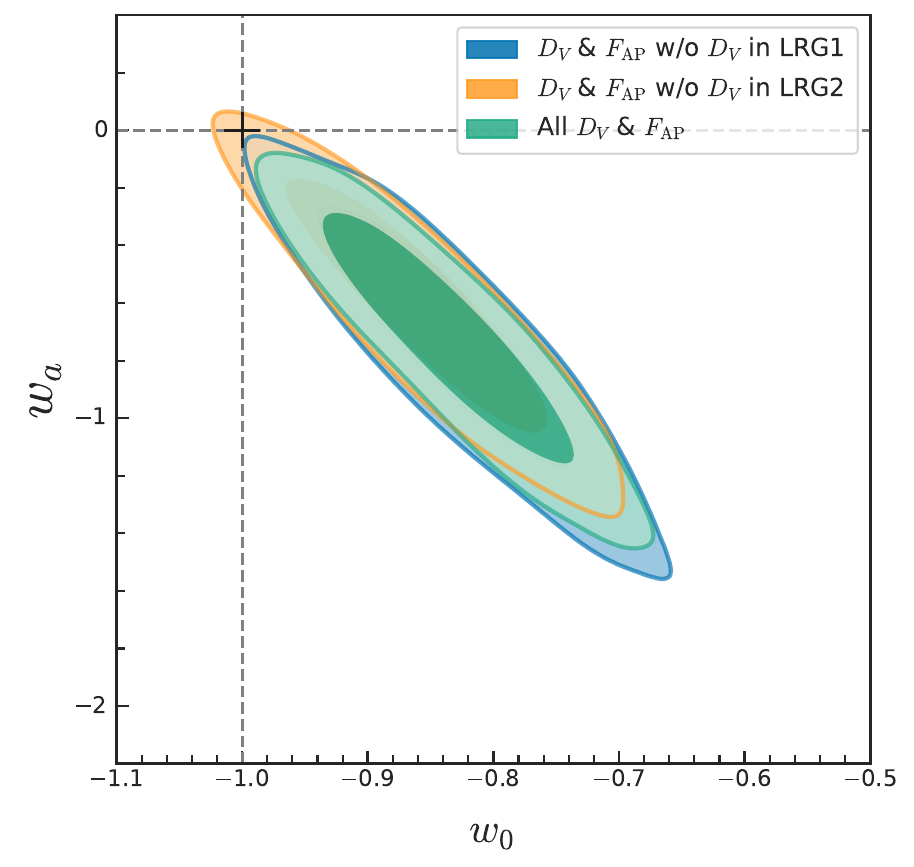}
\includegraphics[height=0.45\textwidth]{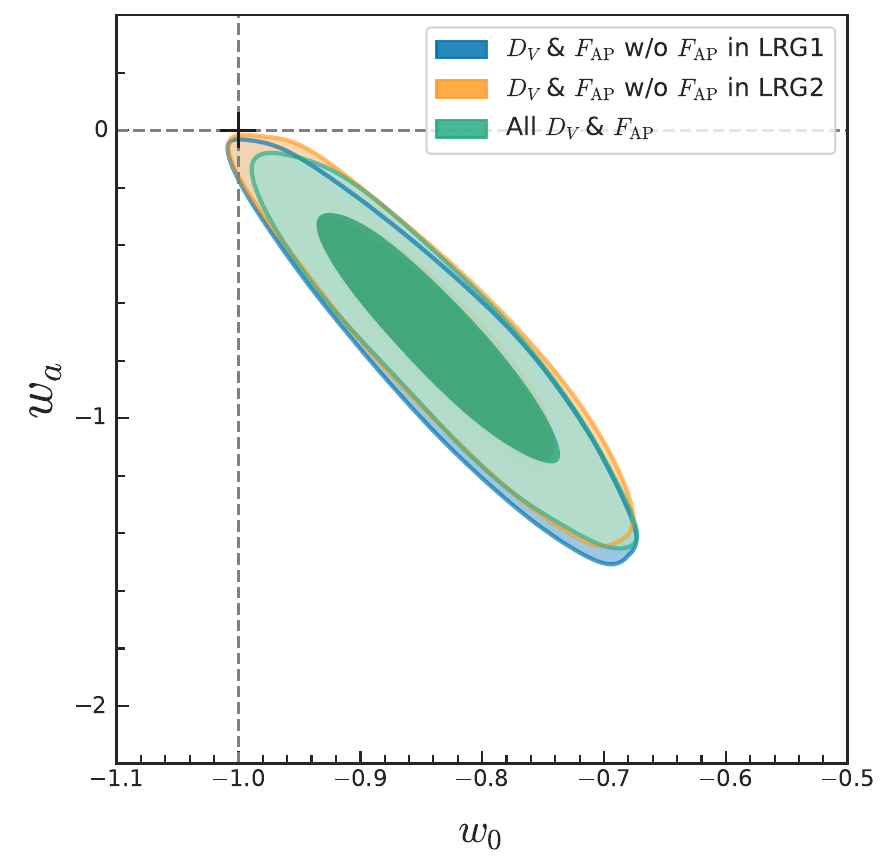}
\caption{\label{fig:w0wa_noDV}\textbf{Left panel}: The marginalized posterior constraint on $w_0$ and $w_a$ by removing $D_V$ in LRG1 (blue), by removing $D_V$ in LRG2 (orange) and by using all the $D_V$ and $F_{\rm AP}$ data (green). 
\textbf{Right panel}: The marginalized posterior constraint on $w_0$ and $w_a$ 
by removing $F_{\rm AP}$ in LRG1 (blue), by removing $F_{\rm AP}$ in LRG2 (orange) and by using all the $D_V$ and $F_{\rm AP}$ data (green).
The results of both panels are from DESI BAO combined with CMB and SNIa. The dark and light shaded region mark $1\sigma (68.3\%)$ and $2\sigma (95.4\%)$ confidence level (CL). The dashed lines and black cross mark the fiducial values of the two parameters in $\Lambda$CDM cosmology.}  
\end{figure}

\begin{figure}[h]
\includegraphics[width=0.5\textwidth]{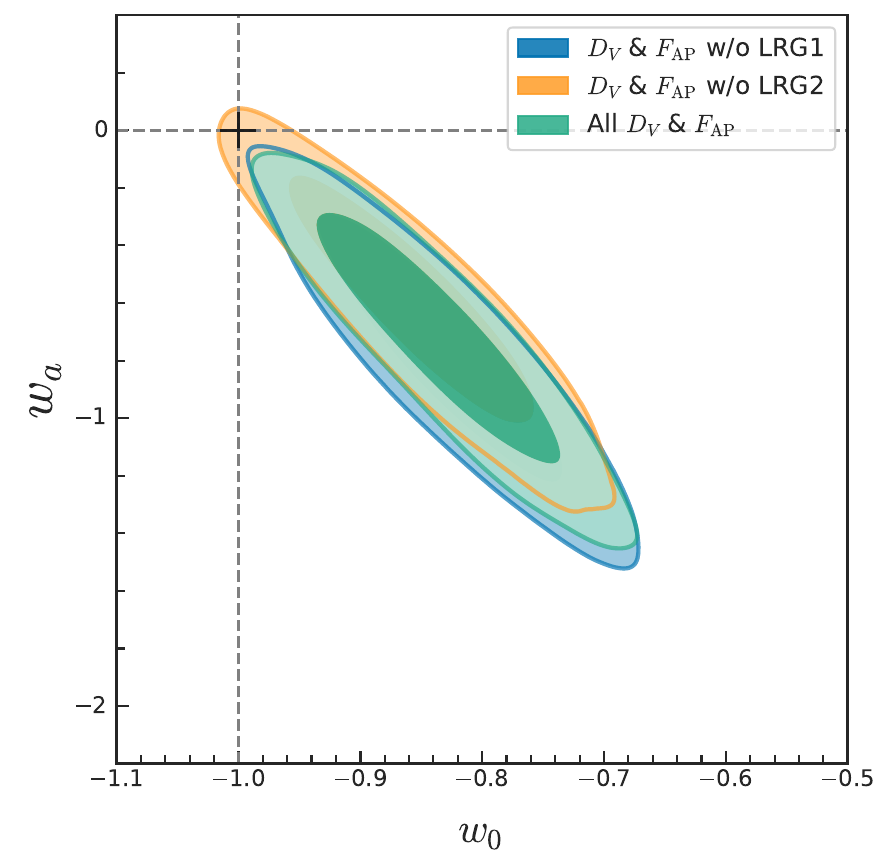}
\caption{\label{fig:w0wa_DVAP} The marginalized posterior constraint on $w_0$ and $w_a$ by removing $D_V$ and $F_{\rm AP}$ in LRG1 (blue), by removing $D_V$ and $F_{\rm AP}$ in LRG2 (orange) and by using all the $D_V$ and $F_{\rm AP}$ data (green). The dark and light shaded region mark $1\sigma (68.3\%)$ and $2\sigma (95.4\%)$ CL, respectively. The dashed lines and black cross mark the fiducial values of the two parameters in $\Lambda$CDM cosmology.}
\end{figure}

\section{Data and Method}\label{Data}
In this section, we first list the data compilation used for this study. In order to validate our result, we adopt the same CMB + BAO + SNIa compilation as the DESI BAO cosmology paper \citep{DESI2024.VI.KP7A}. Then, we present the formulas for the data vector transformation from $(D_M,D_H)$ to $(D_V,F_{\rm AP})$. The mean values and the covariance matrix of the new data vector are provided in this section.   

\begin{itemize}
    \item \textbf{CMB}\; We adopt the Planck 2018 data including the low-$\ell$ temperature (TT), polarization (EE) at multipoles $\ell<30$, Planck (\verb|CamSpec|) high-$\ell$ TTTEEE likelihood \citep{camspec1,camspec2}, and  Atacama Cosmology Telescope (ACT) DR6 CMB lensing likelihood \citep{ACTDR6:1} at multipoles $2\leq\ell\leq4000$.
    
    \item \textbf{BAO}\; We adopt the DESI DR1 BAO data derived from the bright galaxy sample (BGS), LRG, emission line galaxy (ELG), quasar and Lyman-$\alpha$ forest tracers. 
    
    \item \textbf{SN}\; We utilise the SNIa data from the Pantheon+ sample \citep{Pantheon+2022}.
\end{itemize}
The sound horizon at the end of baryon drag epoch leaves an imprint in the distribution of matter, influencing the overdensity of galaxies, thus serving as a standard ruler \citep{Peebles1970,Sunyaev1970}. Measuring 2-point correlation function with position separation vector along the transverse direction, we can obtain the preferred angular separation $\Delta \theta = r_s/D_M $
. Measuring position separation vector along the line-of-sight, we can obtain the preferred redshift separation $\Delta z=r_sH(z)/c=r_s/D_H$. 
We notice that $D_M/r_s$ and $D_H/r_s$ are highly correlated and have similar signal-to-noise ratio, hence, we adopt the spherically-averaged distance $D_V$ to isolate isotropic and anisotropic signals. We follow the same definition of $D_V$ and AP factor $F_{\rm AP}$ as in Ref. \citep{2024arXiv240403002D}, {\it i.e.}
\begin{equation}
D_V=(zD_M^2D_H)^{1/3},\;F_{\rm AP}=\dfrac{D_M}{D_H}.
\end{equation}
For the brevity in the following equations, we define the following terms
\begin{equation}
    d_M=\dfrac{D_M}{r_s},\;d_H=\dfrac{D_H}{r_s},\;d_V=\dfrac{D_V}{r_s}=(zd_M^2d_H)^{1/3},\;F_{\rm AP}=\dfrac{d_M}{d_H}.
\end{equation}
According to the propagation of uncertainties, we can calculate the covariance matrix of $d_V$ and $F_{\rm AP}$ from the one of $d_M$ and $d_H$ by the following equation
\begin{equation}
\label{eq:newcov}
\mathbb{C}_{\alpha\beta}(d_V,F_{\rm AP})=\mathcal{J}_{\alpha i}\mathbb{C}_{ij}(d_M,d_H)\mathcal{J}^{T}_{\beta j},
\end{equation}
where $\mathbb{C}$ denotes the covariance matrix, and $\mathcal{J}$ is the Jacobian matrix defined as
\begin{equation}
 \mathcal{J}=\left[
    \begin{array}{cc}
        \dfrac{\partial d_V}{\partial d_M}&\dfrac{\partial d_V}{\partial d_H}\\
        \vspace{1pt}\\
        \dfrac{\partial F_{\rm AP}}{\partial d_M}&\dfrac{\partial F_{\rm AP}}{\partial d_H}
    \end{array}
\right]   \;.
\end{equation}

Based on Table 1 of Ref. \citep{DESI2024.VI.KP7A}, we have the mean value and covariance matrices of $D_M/r_s$ and $D_H/r_s$ of different tracers. Therefore, we can obtain the mean value, standard deviation, and correlation coefficient of $D_V/r_s$ and $F_{\rm AP}$, via Eq. (\ref{eq:newcov}). We show the result in Table \ref{tab:table1}.

\begin{table}[h]
\begin{ruledtabular}
\begin{tabular}{ccccc}
\textrm{Tracer}&
\textrm{$z_{\rm eff}$}&
\textrm{$D_V/r_s$}&
\textrm{$F_{\rm AP}$}&
\textrm{$r_{\rm off}$}\\
\colrule
\\
\verb|BGS|$^{*}$& 0.295 &  7.93$\pm$0.151 & -               & -     \\
\\
\verb|LRG1|     & 0.510 & 12.57$\pm$0.149 & 0.65$\pm$0.0265 & 0.0532\\
\\
\verb|LRG2|     & 0.706 & 15.90$\pm$0.196 & 0.84$\pm$0.0347 & 0.0495\\
\\
\verb|LRG3+ELG1|& 0.930 & 19.86$\pm$0.170 & 1.21$\pm$0.0330 & 0.0864\\
\\
\verb|ELG2|     & 1.317 & 24.13$\pm$0.365 & 2.01$\pm$0.0948 & 0.2987\\
\\
\verb|QSO|$^{*}$& 1.491 & 26.07$\pm$0.669 & -               & -     \\
\\
${\rm Ly}\alpha$ & 2.330 & 31.52$\pm$0.439 & 4.66$\pm$0.1756 & 0.6052 \\
\\
\end{tabular}
\end{ruledtabular}
\caption{\label{tab:table1}%
The mean value and standard deviation of $d_V$ and $F_{\rm AP}$. $r_{\rm off}=\sigma_{d_V,F_{\rm AP}}/\sqrt{\sigma_{d_V}\sigma_{F_{\rm AP}}}$ is the correlation coefficient between $d_V$ and $F_{\rm AP}$. The superscript `$\ast$' denotes only $d_V$ measured for the tracer.
}
\end{table}

We use the Boltzmann code \verb|camb| \citep{camb2000} and the Bayesian analysis code \verb|cobaya| \citep{Cobaya2019,Cobaya2021} to present the cosmological constraint results obtained from the Monte-Carlo Markov Chain (MCMC) method. 
In this study, we focus on the $w_0w_a$CDM cosmology. We analyze the MCMC chains using \verb|getdist| \citep{Lewis2019}. We adopt the convergence diagnostic of the MCMC chains with the Gelman-Rubin \citep{Gelman_Rubin_1992} criterion $R - 1 \leq 0.01$. For each constraint, we execute 6 chains, each comprising $\sim$ 40,000 steps, and remove 70\% of the burn-in ($\sim$ 72,000 samples remaining). We take the following uniform priors for model parameters: the baryon fraction $\Omega_bh^2 \in [0.005, 0.1]$, the cold dark matter fraction $\Omega_ch^2 \in [0.001, 0.99]$, assuming normal neutrino mass hierarchy $\sum m_\nu = 0.06 {\rm eV}$, the Hubble parameter $H_0\in[20,100]$, the acoustic angular scale at the recombination epoch $100\theta_{\rm MC} \in [0.5, 10]$, the amplitude of primordial power spectrum $\ln(10^{10}A_s) \in [2, 4]$, the primordial scalar spectral index $n_s \in [0.8, 1.2]$, and the reionization optical depth $\tau \in [0.01, 0.8]$.

\begin{table}[h]
\begin{ruledtabular}
\begin{tabular}{lll}
Data &\quad {\boldmath$w_0$} &\quad {\boldmath$w_a$} \\
\hline
\\
All $D_V$ \& $F_{\rm AP}$ & $-0.834\pm 0.064$ & $-0.73^{+0.31}_{-0.27}$ \\
\\
w/o $D_V$ in LRG1 & $-0.835^{+0.060}_{-0.071}$ & $-0.73^{+0.33}_{-0.26}$ \\
\\
w/o $D_V$ in LRG2 & $-0.857\pm 0.067$ & $-0.61\pm 0.29$ \\
\\
w/o $F_{\rm AP}$ in LRG1 & $-0.834\pm 0.067$ & $-0.76\pm 0.30$ \\
\\
w/o $F_{\rm AP}$ in LRG2 & $-0.835\pm 0.066$ & $-0.72\pm 0.29$ \\
\\
w/o LRG1 & $-0.831\pm 0.065$ & $-0.77^{+0.31}_{-0.28}$ \\
\\
w/o LRG2 & $-0.853\pm 0.066$ & $-0.61\pm 0.29$ \\
\end{tabular}
\end{ruledtabular}
\caption{\label{tab:table2}%
The marginalized posterior constraints on $w_0$ and $w_a$ with $1\sigma\; (68.3\%)$ C.L. are shown. In the ``Data'' column, ``w/o $D_V$ in LRG1'' means that we employ $D_V$ \& $F_{\rm AP}$ of all tracers but exclude the data point of $D_V$ in LRG1. Similarly, ``w/o $F_{\rm AP}$ in LRG1'' means that we employ $D_V$ \& $F_{\rm AP}$ of all tracers but not the $F_{\rm AP}$ in LRG1. ``w/o LRG1'' means that we employ all $D_V$ \& $F_{\rm AP}$ but exclude both $D_V$ and $F_{\rm AP}$ in LRG1.
}
\end{table}

\section{Results}\label{results}


In the left panel of Figure \ref{fig:w0wa_noDV}, we show the constraints on $w_0/w_a$ parameters by excluding the $D_V$ data in either LRG1 (blue) or LRG2 (orange). 
The baseline setup, encompassing all $D_V$ and $F_{\rm AP}$ data points, is depicted in green. It is apparent that excluding the $D_V$ data points in LRG1 or LRG2 shifts the contours in the $w_0/w_a$ plane towards the $w=-1$ point (identified by a bold black cross). The distinction lies in the fact that the ``w/o LRG2'' $2\sigma$ contour encompasses the $w=-1$ point, whereas the ``w/o LRG1'' contour does not. 
The right panel of Figure \ref{fig:w0wa_noDV} shows the results by removing $F_{\rm AP}$ data in either LRG1 (blue) or LRG2 (orange). One can see that the overall trend is similar to the one for $D_V$, but with a smaller size of the shift.  
Regardless of whether the $F_{\rm AP}$ in LRG1 or LRG2 is excluded, the resulting $2\sigma$ contours do not intersect with the $w=-1$ point.

In Figure \ref{fig:w0wa_DVAP}, we show the result of simultaneously removing $D_V$ and $F_{\rm AP}$ in either LRG1 or LRG2. One can see that, the result of ``w/o LRG2'' is still consistent with $w=-1$ within $2\sigma$ confidence level; while the result of ``w/o LRG1'' are getting more closer to the one with ``all'' $D_V$ and $F_{\rm AP}$ data points. 
This occurs because the $D_V$ and $F_{\rm AP}$ in LRG1 pull the blue contour in slightly different directions (see Figure \ref{fig:w0wa_noDV}). By incorporating both these data points simultaneously, their effects cancel each other out, resulting in the contour remaining unchanged. This finding aligns the conclusion reported by Wang \citep{2024arXiv240413833W} and the DESI collaboration \citep{DESI2024.VI.KP7A}. We summarise the marginalized posterior constraints on $w_0$ and $w_a$ for various data compilations in Table \ref{tab:table2}. 

\begin{figure}[h]
\includegraphics[width=0.7\textwidth]{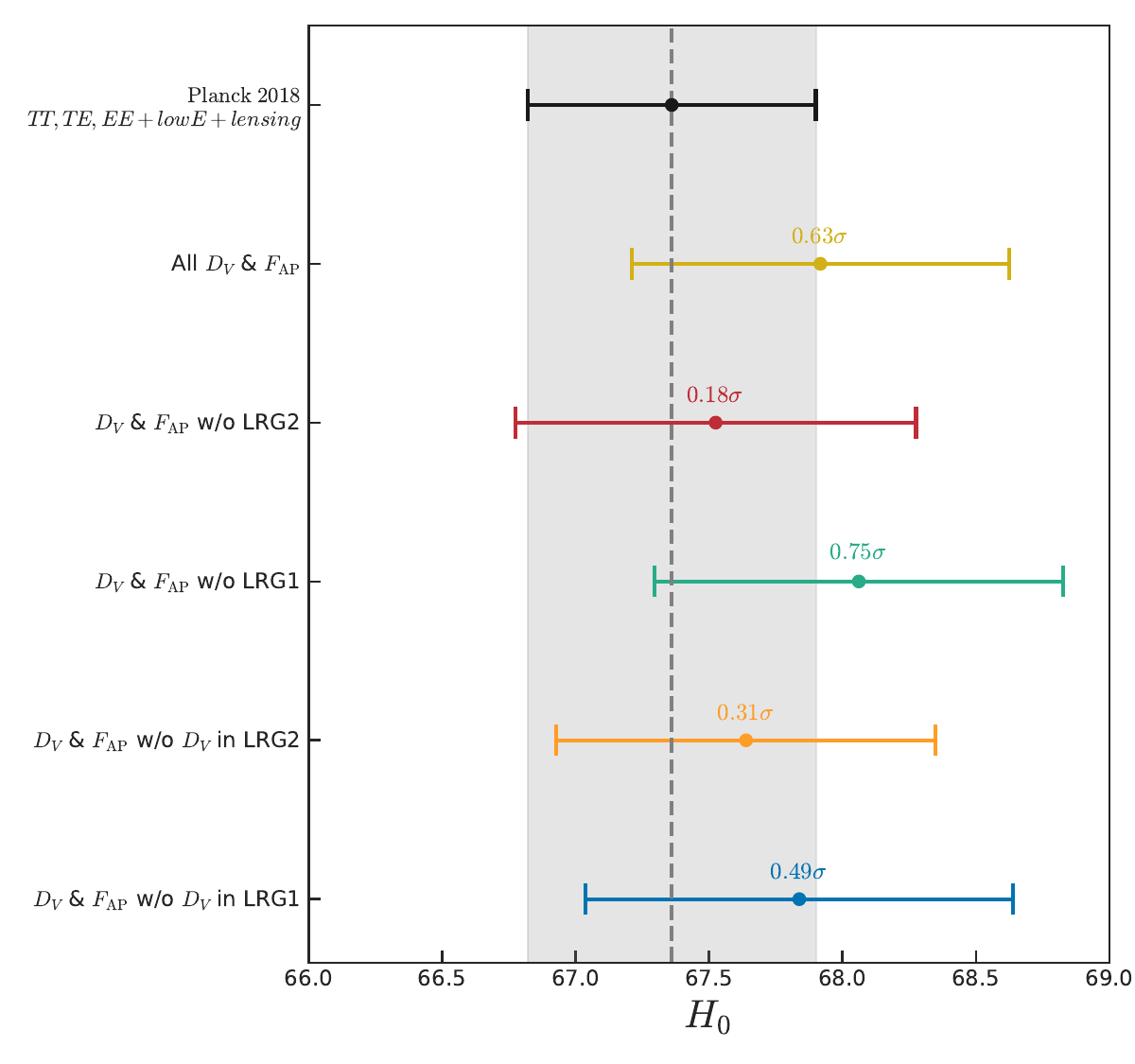}
\caption{\label{fig:H0} The marginalized posterior constraint on $H_0$ from DESI and Planck. The colored numbers indicate the degree of tension between DESI and Planck, measured in units of standard deviation.} 
\end{figure}

In Figure \ref{fig:H0}, we present the $H_0$ constraint results obtained by excluding various data points from DESI. 
If we remove a single $D_V$ data point from either LRG1 or LRG2, the resulting $H_0$ estimate aligns more closely with the Planck result when the $D_V$ data point is removed from LRG2 (orange) compared to LRG1 (blue). Additionally, if we also remove $F_{\rm AP}$ from the corresponding data bin, the $H_0$ estimate for LRG1 shifts even closer to the Planck result (red), while the estimate for LRG2 moves further away (green).

To safeguard the integrity of our findings against potential influences from factors such as inaccuracies in data vector transformation or convergence issues, we validate our results in both the $(D_M, D_H)$ base and the $(D_V, F_{\rm AP})$ base (refer to Figure \ref{fig:w0wa_DMDH_DVAP} and Figure \ref{fig:w0wa_nolrg12} in Appendix \ref{validation}).
One can see that the two results perfectly align on top of each other.


For the first time, DESI BAO shows the hint of dynamical dark energy. This is an exciting result. In the meantime, we should remain cautious regarding the data. 
The BAO distance is derived from fitting the monopole and quadrupole components of galaxy clustering. Notably, we have observed unexpected discrepancies in both the monopole and quadrupole between the "true" cosmology and the fiducial cosmology in LRG1 and LRG2 datasets. This observation prompts us to investigate which observable serves as the primary driver behind this inconsistency. To achieve this objective, we convert the official $(D_M,D_H)$ data vector to $(D_V,F_{\rm AP})$. This is because the latter is more directly linked to the monopole and quadrupole components in the galaxy clustering. Through a step-by-step exclusion of data points, we ascertain that the angle-averaged distance $D_V$ in LRG2 predominantly favors the presence of dynamical dark energy.
This suggests that we should allocate greater attention to the monopole component in LRG2 in the future DESI three-year and five-year analysis.

\begin{acknowledgments}
ZW, SL and BH acknowledge the science research grants from the China Manned Space Project with No. CMS-CSST-2021-A12. ZD acknowledges the grant from the National Science Foundation with No. 12273020.
\end{acknowledgments}

\bibliography{sorsamp}

\appendix
\section{Validation}\label{validation}
\begin{figure}[h]
\includegraphics[width=0.5\textwidth]{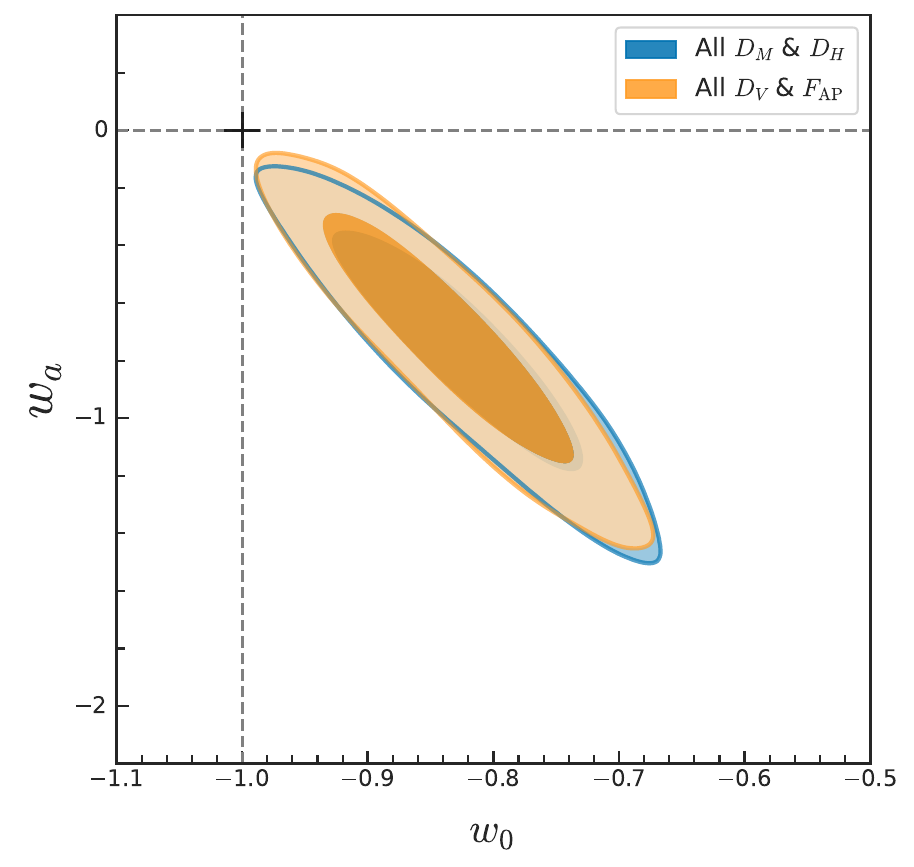}
\caption{\label{fig:w0wa_DMDH_DVAP} The marginalized posterior constraint on $w_0$ and $w_a$ comparison from all tracers of DESI BAO employing $(D_M, D_H)$ (blue) and $(D_V, F_{\rm AP})$ (orange).  The dark and light shaded region mark $1\sigma (68.3\%)$ and $2\sigma (95.4\%)$ CL.}  
\end{figure}

\begin{figure}[h]
\includegraphics[height=0.45\textwidth]{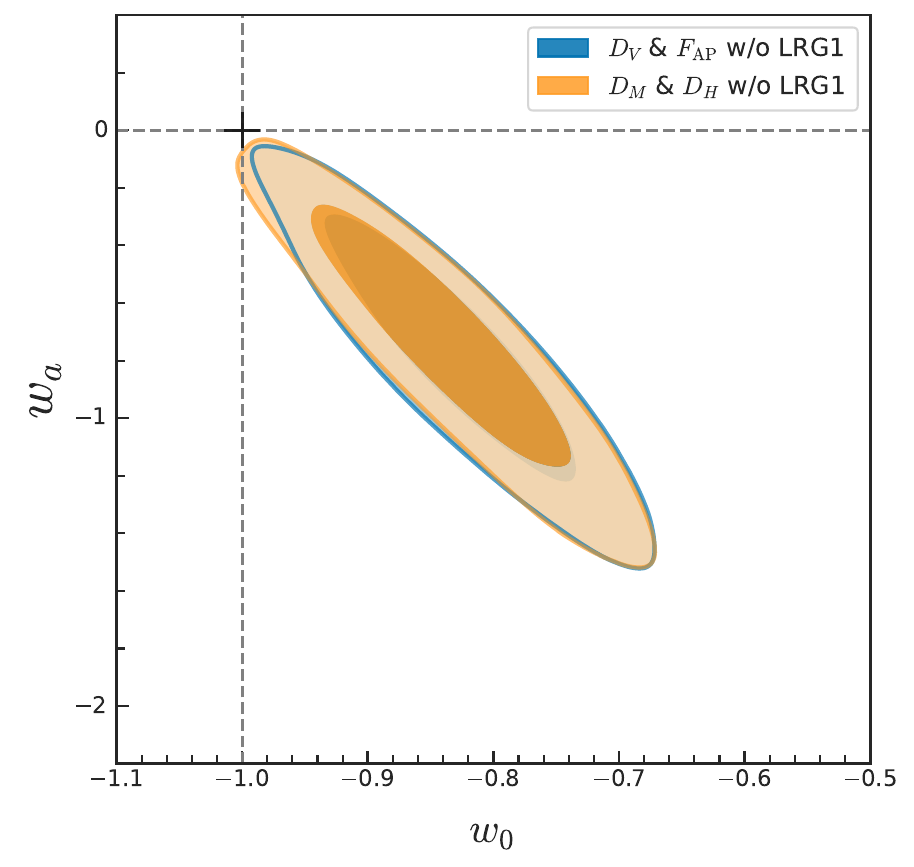}
\includegraphics[height=0.45\textwidth]{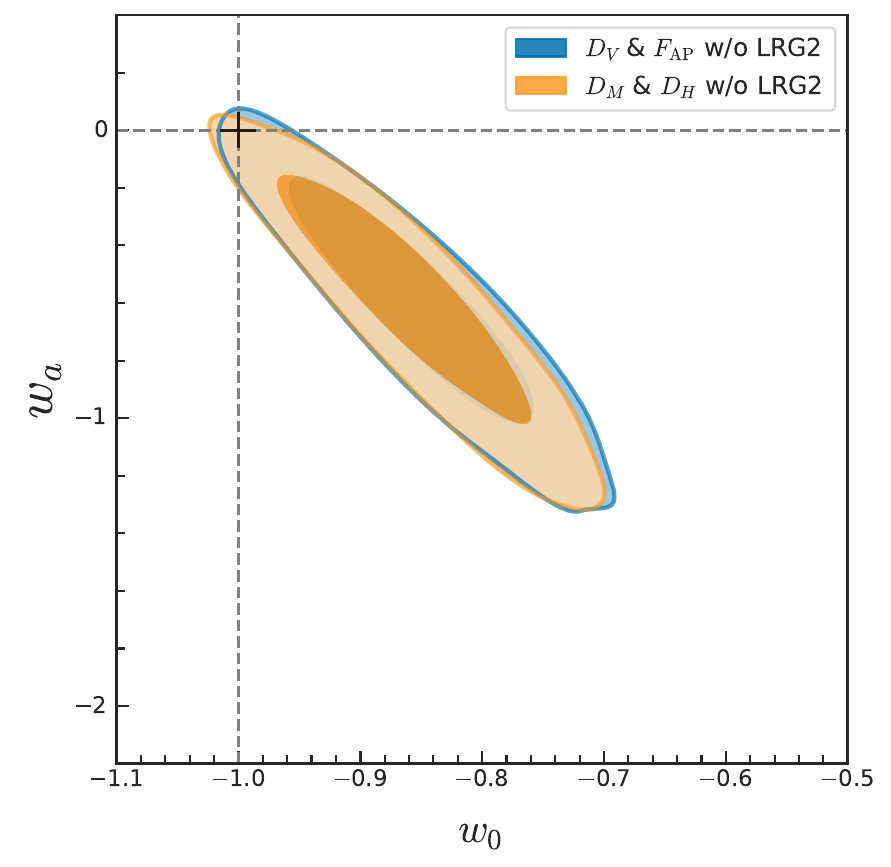}
\caption{\label{fig:w0wa_nolrg12}\textbf{Left panel}: Comparison of the marginalized posterior constraint on $w_0$ and $w_a$ from $(D_V, F_{\rm AP})$ (blue) versus $(D_M, D_H)$ (orange) with all tracers but not LRG1. \textbf{Right panel}: Comparison of the marginalized posterior constraint on $w_0$ and $w_a$ from $(D_V, F_{\rm AP})$ (blue) versus $(D_M, D_H)$ (orange) with all tracers but not LRG2. The dark and light shaded region mark $1\sigma (68.3\%)$ and $2\sigma (95.4\%)$ CL.
}
\end{figure}

In order to examine the consistency of parameter constraints, particularly concerning the efficacy of $(D_M, D_H)$ vs. $(D_V, F_{\rm AP})$ in constraining $w_0$ and $w_a$, we present the consistency test from $(D_M, D_H)$ versus $(D_V, F_{\rm AP})$ with all tracers in Figure~\ref{fig:w0wa_DMDH_DVAP}. We also check the agreement of the constraint from $(D_M, D_H)$ versus $(D_V, F_{\rm AP})$ with LRG1 or LRG2 excluded, as shown in Figure~\ref{fig:w0wa_nolrg12}. 


\end{document}